\documentclass[preprintnumbers,amsmath,amssymb]{revtex4}
\usepackage{epsfig}
\usepackage{color}
\usepackage{dsfont}
\usepackage[colorlinks,urlcolor=blue,linkcolor=blue,citecolor=red]{hyperref}
\begin{document}
\setlength{\voffset}{1.0cm}
\title{Exact tricritical point from next-to-leading-order stability analysis}
\author{Michael Thies\footnote{michael.thies@gravity.fau.de}}
\affiliation{Institut f\"ur  Theoretische Physik, Universit\"at Erlangen-N\"urnberg, D-91058, Erlangen, Germany}
\date{\today}

\begin{abstract}
In the massive chiral Gross-Neveu model, a phase boundary separates a homogeneous from an inhomogeneous phase. It 
consists of two parts, a second order line and a first order line, joined at a tricritical point. Whereas the first order phase
boundary requires a full, numerical Hartree-Fock calculation, the second order phase boundary can be determined exactly 
and with less effort by a perturbative stability analysis. We extend this stability analysis to higher order perturbation theory.
This enables us to locate the tricritical point exactly, without need to perform a Hartree-Fock calculation. Divergencies
due to the emergence of  spectral gaps in a spatially periodic perturbation are handled using well established tools from many body
theory.

\end{abstract}

%\pacs{}
\maketitle

%<<<<<<<<<<<<<<<<<<<<<<<<<<<<<<<<<<<<<<<<<<<<<<<<<<<<<<<<<<<<<<<<<<<<<<<<<<<<<<<<<<<<<<<<<<<< <<<<<<<<<<<<<<<<<<<<<<<<<<<<<
%<<<<<<<<<<<<<<<<<<<<<<<<<<<<<<<<<<<<<<<<<<<<<<<<<<<<<<<<<<<<<<<<<<<<<<<<<<<<<<<<<<<<<<<<<<<<<<<<<<<<<<<<<<<<<<<<<<<<<<<<<<
\section{Introduction}
\label{sect1}
%<<<<<<<<<<<<<<<<<<<<<<<<<<<<<<<<<<<<<<<<<<<<<<<<<<<<<<<<<<<<<<<<<<<<<<<<<<<<<<<<<<<<<<<<<<<<<<<<<<<<<<<<<<<<<<<<<<<<<<<<<<
%<<<<<<<<<<<<<<<<<<<<<<<<<<<<<<<<<<<<<<<<<<<<<<<<<<<<<<<<<<<<<<<<<<<<<<<<<<<<<<<<<<<<<<<<<<<<<<<<<<<<<<<<<<<<<<<<<<<<<<<<<<
The present paper is about the phase diagram of the massive chiral Gross-Neveu ($\chi$GN) model \cite{L1}. This model can be regarded as a 1+1 dimensional
version of the Nambu--Jona-Lasinio (NJL) model \cite{L2} with U(1)$\times$U(1) chiral symmetry, explicitly broken by a bare mass term. The Lagrangian reads
\begin{equation}
{\cal L} = \bar{\psi} (i \partial\!\!\!/- m_b) \psi + \frac{g^2}{2} \left[ (\bar{\psi}\psi)^2 + (\bar{\psi} i \gamma_5 \psi)^2\right].
\label{1.1}
\end{equation} 
Flavor indices are suppressed as usual, and we are working in the 't Hooft limit \cite{L3} ($N \to \infty, Ng^2=$ const.) with semiclassical methods.

The phase diagrams of GN type models have proven to be quite instructive. On the one hand, they have led to exact results which could be used 
as a testing ground for new numerical methods \cite{L4,L5} and other ideas \cite{L6,L7} related to quantum chromodynamics (QCD). On the other hand, by 
exhibiting a variety of inhomogeneous phases in some regions of the phase diagram, they have triggered some activity for looking for
similar phenomena in higher dimensional theories \cite{L8,L9,L10}. This interest has been reinforced recently by the observation that some of the results
do not seem to be an artefact of the large $N$ limit, but leave their traces in models with a finite number of flavors \cite{L11,L12}, even down to $N=2$ \cite{L13,L14}.
From a theoretical point of view, model (\ref{1.1}) is more challenging than the massive GN model with discrete chiral symmetry \cite{L15}.
With the exception of the chiral limit with its ``chiral spiral" phase \cite{L16,L17}, it does not seem possible to study the full phase diagram of the
$\chi$GN model analytically. We think that such studies are nevertheless worthwhile to guide our intuition and help us develop techniques
useful in more realistic situations.

%###########################################################################################################################
\begin{figure}
\begin{center}
\epsfig{file=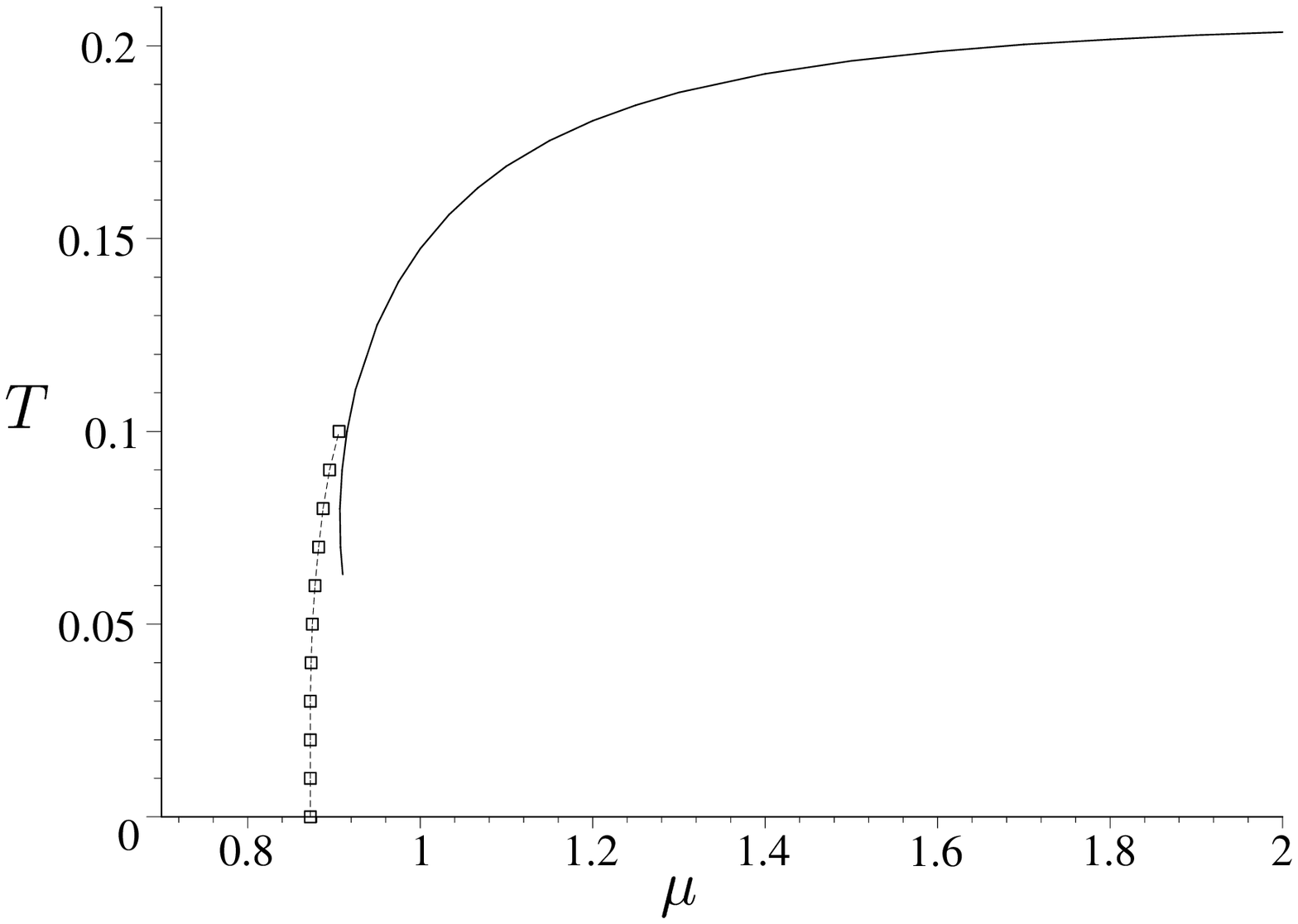,height=8cm,width=6cm,angle=270}
\caption{Phase boundary between homogeneous and inhomogeneous phases of massive $\chi$GN model at $\gamma=1.0$ \cite{L18}. Solid line: second order transition, stability
analysis. Dots: First order transition, Hartree-Fock calculation. The two branches should meet at a tricritical point.}
\label{fig1}
\end{center}
\end{figure}
%####################################################################################################\#######################
To explain our goal, we show in Fig.~\ref{fig1} a typical phase boundary for one particular bare fermion mass \cite{L18}. To the right and below the curve,
there is an inhomogeneous crystal phase. To the left and above the curve, the system is in a homogeneous phase of massive fermions.   
The phase boundary comprises two distinct parts. The solid line corresponds to a 2nd order transition. When crossing it, the homogeneous phase becomes
unstable against developing a spatial oscillation with finite wave number and infinitesimal amplitude. This part can be computed rather easily by a 
straightforward stability analysis, based on a leading order (LO) perturbative treatment of the inhomogeneous mean field. The dots belong to a 
first order phase boundary. When crossing this boundary, the system jumps from the homogeneous phase to a crystal phase with finite wave number and finite
amplitude. Here perturbation theory (PT) cannot be used, but one needs to do a full, numerical Hartree-Fock (HF) calculation and find the points where 
two distinct solutions are degenerate. These two parts of the phase boundary must be joined at a tricritical point not shown in Fig.~\ref{fig1}. Here, the situation is somewhat
frustrating. If one approaches the tricritical point following the 2nd order phase boundary (``top down"), there is no signal whatsoever and one simply crosses the tricritical point,
continuing into an unphysical curve. If one approaches it from the 1st order phase boundary (``bottom up"), computations get more and more delicate
since one has to minimize a function with several nearby, shallow, almost degenerate minima. This is actually the way the tricritical point 
has been determined in Ref. \cite{L18}, using some extrapolation of unknown accuracy.  

The LO stability analysis underlying the solid line is a standard tool in such investigations. Several recent studies are dedicated to this type of stability analysis,
using alternative techniques and trying to extend its range of applicability \cite{L19,L20}. In our case, it has proven to be an efficient way of determining the exact phase boundary. The fact that
we call it perturbative does not mean that it is completely trivial though. Indeed, it is well known that naive PT breaks down near the gaps
generated by a periodic potential. This difficulty can be overcome by ``almost degenerate perturbation theory" (ADPT) \cite{L21}. 

The tricritical point lies
on the perturbative phase boundary, so that only one additional condition is needed to pin it down. It is hard to believe that the only way out  is
a full HF calculation, minimizing a large number of parameters in a region where it is most difficult. In this work, we will try to formulate this missing condition 
and work out a ``top down" approach to the tricritical point, independent of the HF calculation and hopefully exact.  In the same way as one can characterize
stationary points of an ordinary function only by looking at the 2nd derivative, it is clear that we have to extend PT to next-to-leading-order (NLO).
This raises immediately questions about divergencies in higher order PT with periodic potentials which need to be addressed here.

Everything we will discuss is directly applicable to a generalized version of the $\chi$GN model with isospin, where the pseudoscalar interaction term
is replaced by $(\bar{\psi} i\gamma_5 \vec{\tau}\psi)^2$ and chiral symmetry gets promoted to the group SU(2)$\times$SU(2) \cite{L22,L23,L24}. Here, nothing is 
known yet about the tricritical points. This part of our study will be left to a forthcoming paper. 

This paper is organized as follows: In Sect.~\ref{sect2} we reconsider the Ginzburg-Landau (GL) approach in a region where the condensates are weak and slowly varying.
This ``warm-up" problem is useful for understanding how to find the tricritical point using a perturbative approach. Sect.~\ref{sect3} deals with the spectrum of 
massive fermions, perturbed by a spatially periodic potential. The main focus will be on the issue how to avoid divergencies arising at the spectral gaps
in this type of potential. Sect.~\ref{sect4} is the central part of this investigation, showing how to locate the tricritical point without a full HF calculation.
Our results will be presented both in tabular form and in figures and compared to previous results. We end with a short summary and conclusions, Sect.~\ref{sect5}.

%<<<<<<<<<<<<<<<<<<<<<<<<<<<<<<<<<<<<<<<<<<<<<<<<<<<<<<<<<<<<<<<<<<<<<<<<<<<<<<<<<<<<<<<<<<<< <<<<<<<<<<<<<<<<<<<<<<<<<<<<<
%<<<<<<<<<<<<<<<<<<<<<<<<<<<<<<<<<<<<<<<<<<<<<<<<<<<<<<<<<<<<<<<<<<<<<<<<<<<<<<<<<<<<<<<<<<<<<<<<<<<<<<<<<<<<<<<<<<<<<<<<<<
\section{Ginzburg-Landau approach}
\label{sect2}
%<<<<<<<<<<<<<<<<<<<<<<<<<<<<<<<<<<<<<<<<<<<<<<<<<<<<<<<<<<<<<<<<<<<<<<<<<<<<<<<<<<<<<<<<<<<<<<<<<<<<<<<<<<<<<<<<<<<<<<<<<<
%<<<<<<<<<<<<<<<<<<<<<<<<<<<<<<<<<<<<<<<<<<<<<<<<<<<<<<<<<<<<<<<<<<<<<<<<<<<<<<<<<<<<<<<<<<<<<<<<<<<<<<<<<<<<<<<<<<<<<<<<<<
We first recall some basic facts about the (large $N$) $\chi$GN model in 1+1 dimensions. In the chiral limit, a semiclassical analysis shows that the fermions
acquire a mass dynamically in the vacuum. If one heats up the system, the mass decreases until it vanishes at a critical temperature $T_c=e^{\rm C}/\pi\approx 0.567$
(C is the Euler constant). If one now switches on a bare fermion mass (confinement parameter $\gamma=\pi m_b/Ng^2$) and the chemical potential $\mu$, there is a region
in ($\gamma,\mu,T$) space around the point($0,0, T_c$) where scalar and pseudoscalar condensates are both weak and slowly varying. As discussed in
Ref. \cite{L25}, this region is accessible via a Ginzburg-Landau (GL) approach derived from a gradient expansion \cite{L26} of the grand canonical potential. The GL effective
action density can be expressed in terms of a complex scalar field $\Phi=S-iP$ as
\begin{equation}
\Psi_{\rm eff}  =   \alpha_0 + \alpha_1 \left( |\Phi|^2-2 {\rm Re\,} \Phi\right) + \alpha_2 |\Phi|^2
 + \alpha_3 {\rm Im\,} \Phi (\Phi')^* + \alpha_4 \left( |\Phi|^4+ |\Phi'|^2\right).
\label{2.1}
\end{equation}
We have only written down the terms up to the order at which one first ``sees" the tricritical point.
The coefficients $\alpha_n$ are analytically calculable functions of ($\mu,T$). 
In the region where
\begin{equation}
\gamma \sim \epsilon^3, \quad \mu \sim \epsilon, \quad \tau =\sqrt{T_c-T} \sim \epsilon,
\label{2.2}
\end{equation}
they can be approximated by
\begin{eqnarray}
\alpha_1 & = & \frac{\gamma}{2\pi},
\nonumber \\
\alpha_2 & = & \frac{2}{\pi a} \mu^2 - \frac{1}{2\pi T_c} \tau^2 + {\rm O}(\epsilon^4),
\nonumber \\
\alpha_3 & = &  \frac{2}{\pi a} \mu  + {\rm O}(\epsilon^3),
\nonumber \\
\alpha_4 & = & \frac{1}{2\pi a}  + {\rm O}(\epsilon^2),
\label{2.3}
\end{eqnarray}
with  
\begin{equation}
a = \frac{16 e^{2{\rm C}}}{7 \zeta(3)} \approx 6.032 
\label{2.4}
\end{equation}
By introducing a rescaled scalar field $\varphi$ through
\begin{equation}
\Phi(x) = \gamma^{1/3} \varphi(\xi), \quad \xi = \gamma^{1/3} x
\label{2.5}
\end{equation}
as well as rescaled parameters for chemical potential and temperature,
\begin{equation}
\nu = 2 \gamma^{-1/3} \mu, \quad \sigma = \sqrt{\frac{a}{T_c}} \gamma^{-1/3} \tau,
\label{2.6}
\end{equation}
the $\gamma$ dependence of the effective action disappears up to an overall factor,
\begin{equation}
\Psi_{\rm eff} = \frac{\gamma}{2 \pi a}  \left( -2a {\rm Re\,}\varphi + (\nu^2-\sigma^2) |\varphi|^2 + 2 \nu {\rm Im\,} \varphi (\dot{\varphi})^* + |\varphi|^4+ |\dot{\varphi}|^2 \right).
\label{2.7}
\end{equation}
(A dot denotes the derivative with respect to $\xi$). This behavior under scaling allows one to draw a ``universal" phase boundary independent of $\gamma$ in the $(\nu,\sigma$) plane,
as opposed to a family of strongly $\gamma$ dependent curves in the $(\mu,T)$ plane. 

This GL model was already largely solved in Ref. \cite{L25} to which we refer the reader for figures and more details. Here we shall use it as a warm-up to understand
how to locate the exact tricritical curve in ($\gamma,\mu,T$) space. In \cite{L25} the perturbative phase boundary between the homogeneous and inhomogeneous phases  
was found analytically with the help of a stability analysis. By contrast, the non-perturbative, first order phase boundary could only be constructed at the cost of solving the Euler-Lagrange 
equations derived from the effective action (\ref{2.7}) numerically. The tricritical point was then also determined numerically, approaching it from the non-perturbative side (``bottom up"). Here we are looking for
an alternative method for finding the exact tricritical point. The idea is to approach it from the perturbative side (``top down"), thus avoiding numerical computations in a region where they are most 
difficult. We will be careful to develop the method in such a way that it can also be applied to a full HF calculation, beyond the region of validity of GL theory.

A sufficiently general ansatz for a stability analysis in the GL model is
\begin{equation}
\varphi = m+A \cos(q\xi) +i B \sin(q\xi).
\label{2.8}
\end{equation}
Inserting it into (\ref{2.7}) and averaging over one period, we find
\begin{eqnarray}
\Psi_{\rm eff} & = & \frac{\gamma}{2\pi a} \psi_{\rm eff},
\nonumber \\ 
\psi_{\rm eff} & = & \left. \psi_{\rm eff}\right|_{\rm hom} +\left. \psi_{\rm eff}\right|_{\rm inhom},
\nonumber \\
\left. \psi_{\rm eff}\right|_{\rm hom} & = &  -2am+ (\nu^2-\sigma^2) m^2 + m^4,
\nonumber \\
\left. \psi_{\rm eff}\right|_{\rm inhom}
 & = &  \left( 6 m^2 + q^2 +\nu^2-\sigma^2\right)\frac{A^2}{2} + \left( 2 m^2 + q^2 +\nu^2 -\sigma^2\right) \frac{B^2}{2}
\nonumber \\
& &  -2 \nu q A B  + \frac{1}{8} \left( 3 A^4 + 3 B^4 + 2 A^2 B^2 \right) .
\label{2.9}
\end{eqnarray}
Here, $\left. \psi_{\rm eff}\right|_{\rm hom}$ is the 0-th order term whereas quadratic (quartic) contributions to $\left. \psi_{\rm eff}\right|_{\rm inhom}$ in ($A,B$) 
will be referred to as LO (NLO) terms.

We first construct the perturbative phase boundary between homogeneous and inhomogeneous phases
via a LO stability analysis. We need to keep 0-th and LO terms only, 
\begin{equation}
\left. \psi_{\rm eff}\right|_{\rm LO}  =   -2am+ (\nu^2-\sigma^2) m^2 + m^4 +
  \left( 6 m^2 + q^2 +\nu^2-\sigma^2\right)\frac{A^2}{2} + \left( 2 m^2 + q^2 +\nu^2 -\sigma^2\right) \frac{B^2}{2}   -2 \nu q A B   .
\label{2.10}
\end{equation}
This has to be minimized with respect to $m,A,B,q$. Variation with respect to $m$ on the phase boundary ($A=B=0$) is
the same as in the homogeneous calculation,
\begin{equation}
\partial_m \psi_{\rm eff} = 0 = -2a + 2 (\nu^2-\sigma^2)m + 4 m^3 .
\label{2.11}
\end{equation}
Variation with respect to $A,B,q$ yields
\begin{eqnarray}
\partial_A \psi_{\rm eff} & = & 0 = -2\nu B q + A(6 m^2+q^2+\nu^2-\sigma^2),
\label{2.12} \\
\partial_B \psi_{\rm eff} & = & 0 = - 2 \nu A q + B(2 m^2 + q^2 + \nu^2 - \sigma^2),
\label{2.13} \\
\partial_q \psi_{\rm eff} & = & 0 = q(A^2+B^2)-2\nu A B.
\label{2.14}
\end{eqnarray}
Here, $m$ should be identified with the solution of (\ref{2.11}) on the phase boundary.
Mimicking the procedure used in the full HF calculation of the $\chi$GN model, we write the LO inhomogeneous 
action density as the quadratic form 
\begin{equation}
\left. \psi_{\rm eff}\right|_{\rm inhom} = {\cal M}_{11} A^2+ 2 {\cal M}_{12} AB + {\cal M}_{22} B^2 .
\label{2.15}
\end{equation}
The condition for the existence of a nontrivial solution of Eqs.~(\ref{2.12},\ref{2.13}) is 
\begin{equation}
{\rm det\,} {\cal M} = 0.
\label{2.16}
\end{equation}
Besides, the homogeneous linear system yields the ratio
\begin{equation}
R = \frac{A}{B} = - \frac{{\cal M}_{12}}{{\cal M}_{11}}.
\label{2.17}
\end{equation}
Condition (\ref{2.14}) can then be shown to be equivalent to
\begin{equation}
\partial_q {\rm det\,} {\cal M} = 0.
\label{2.18}
\end{equation}
In the full HF calculation, one would determine the perturbative phase boundary as follows: choose a chemical potential and follow ${\rm det\,}{\cal M}$
as a function of $q,T$ up to the point where ${\rm det\,}{\cal M}$ and $\partial_q {\rm det\,}{\cal M}$ vanish simultaneously. This yields the critical temperature 
at this value of $\mu$, the ratio $R$ and the wave number $q$ characterizing
the instability. The GL case is simple enough so that everything can be done analytically. 
Solve Eq.~(\ref{2.13}) for $q$,
\begin{equation}
q = \frac{2\nu R}{R^2+1},
\label{2.19}
\end{equation}
and insert the result into Eqs.~(\ref{2.12},{\ref{2.13}), 
\begin{eqnarray}
R^2 & = & \frac{\nu^2 + \sigma^2 - 4 m^2}{\nu^2 - \sigma^2 + 6 m^2},
\label{2.20} \\
0 & = & 4 m^2 \nu^2 - m^4 - \nu^2 \sigma^2.
\label{2.21}
\end{eqnarray}
Eq.~(\ref{2.19}) then yields
\begin{equation}
q^2 = \frac{\nu^2 (\nu^2+\sigma^2-4m^2)(\nu^2-\sigma^2+6m^2)}{(\nu^2+m^2)^2}.
\label{2.22}
\end{equation}
The preferred strategy would be to solve the mass equation (\ref{2.11}) for $m=m_0$, plug $m_0$ into the other equations and get the phase boundary, $R$ and $q$ as
a function of $\nu,\sigma$.  As this is not possible explicitly, it is better to use $m_0$ as curve parameter of the phase boundary.
To this end, one solves Eqs.~(\ref{2.11}) and (\ref{2.21}) for $\nu^2,\sigma^2$.
This yields a parametric representation of the phase boundary
\begin{eqnarray}
2 m_0 \nu^2 & = &  a + 2 m_0^3 + \sqrt{a(a+4m_0^3)},
\nonumber \\
2 m_0 \sigma^2 & = &  -a + 6 m_0^3 + \sqrt{a(a+4m_0^3)}.
\label{2.23}
\end{eqnarray}
Inserting $\nu^2,\sigma^2$ on the phase boundary into $R^2,q^2$ from Eqs.~(\ref{2.20}) and (\ref{2.22}), we find
\begin{eqnarray}
q^2 & = & \frac{\sqrt{a(a+4m_0^3)}}{m_0},
\nonumber \\
R^2 & = & \frac{a}{\sqrt{a(a+4m_0^3)}}.
\label{2.24}
\end{eqnarray}
This completes the LO stability analysis. We have obtained the curve where the instability sets in, Eq.~(\ref{2.23}), but also some useful information about
the mode responsible for the instability. It is characterized by $R$ and $q$. By contrast, the overall strength parameter $B$ of the perturbation remains
undetermined to LO.

It is known from numerical computations of the phase diagrams of massive $\chi$GN models that the perturbative phase boundary
(2nd order transition) ends at a tricritical point where a non-perturbative, first order transition sets in. 
Although we know that the tricritical point lies on the perturbative phase boundary, there is clearly no way of determining its position to LO.
In Ref. \cite{L25} as well as in previous full HF calculations \cite{L18}, it was determined by pushing the numerical computation of the 1st order phase boundary towards the endpoint.
Even within this GL toy model, there has been no exact determination of the tricritical point so far. The numerical values obtained in Ref. \cite{L25} by extrapolation from the first order side are 
\begin{equation}
m_{\rm tri} \approx 0.78,  \quad \nu_{\rm tri} \approx 2.99, \quad \sigma_{\rm tri} \approx 1.54 
\label{2.25}
\end{equation}
These numbers have been challenged by a full numerical HF solution of the $\chi$GN model, reporting the values $\nu_{\rm tri} \approx 3.039, \sigma_{\rm tri} \approx 1.464$ \cite{L18}.
This discrepancy already indicates difficulties in locating the tricritical point.
Our goal here is to develop an alternative, potentially exact method for determining the tricritical point,
independently of the numerical solution of Euler-Lagrange or HF equations. 

Let us go back to the full effective action, but keeping $q,R$ at the LO values (\ref{2.24}). We are not yet allowed to replace $m$ by $m_0$ though. We push the stability analysis to next order,
but only in the direction of the unstable mode. It is advantageous to use the variable $\omega=B^2$ for this purpose,
\begin{eqnarray}
{\cal V}_{\rm eff} & = & \left. {\cal V}_{\rm eff}\right|_{\rm hom} +\left. {\cal V}_{\rm eff}\right|_{\rm inhom},
\nonumber \\
\left. {\cal V}_{\rm eff}\right|_{\rm hom} & = &  -2a m+ (\nu^2-\sigma^2) m^2 + m^4,
\nonumber \\
\left. {\cal V}_{\rm eff}\right|_{\rm inhom}
 & = &  \left( 6 m^2 + q^2 +\nu^2-\sigma^2\right)\frac{R^2\omega}{2} + \left( 2 m^2 + q^2 +\nu^2 -\sigma^2\right) \frac{\omega}{2}
\nonumber \\
& &  -2 \nu q R \omega  + \frac{1}{8} \left( 3 R^4  + 2 R^2 +3)\omega^2 \right) .
\label{2.26}
\end{eqnarray}
Only two out of the four original variables are left, $m$ and $\omega$. Consider the Hesse matrix
\begin{equation}
H = \left( \begin{array}{cc} \partial_m^2 {\cal V}_{\rm eff} & \partial_m \partial_{\omega} {\cal V}_{\rm eff} \\ \partial_{\omega}\partial_m {\cal V}_{\rm eff} & \partial_{\omega}^2 {\cal V}_{\rm eff} \end{array} \right) .
\label{2.27}
\end{equation}
Taking the required partial derivatives at the point $m=m_0,\omega=0$ on the perturbative phase boundary yields
\begin{equation}
H = \left( \begin{array}{cc} 2 (\nu^2-\sigma^2) + 12 m_0^2 & 2m_0(3 R^2+1 ) \\ 2 m_0(3 R^2+1) & \frac{1}{4}(3 R^4+2 R^2+1) \end{array} \right)
\label{2.28}
\end{equation}
Notice that $H_{11}$ depends only on the 0-th order, $H_{12}$ only on the LO effective action. $H_{22}$ is the only entry requiring the NLO effective action.
The tricritical point can now be found by demanding that the Hesse determinant vanishes, 
\begin{equation}
{\rm det\,}H = 0 = R^2(3 R^2+2)(20 m_0^3-a)-4 m_0^3-3a ,
\label{2.29}
\end{equation}
where we have used $\nu^2,\sigma^2$ from (\ref{2.23}). Inserting our result for $R$ (\ref{2.24}) and expanding the 
equation such that the square roots disappear leads to the quartic equation
\begin{equation}
0 = 16 z^4 - 488 z^3 +73 z^2  -24 z +2, \quad  z=\frac{1}{a} m_{\rm tri}^3.
\label{2.30}
\end{equation}
We have to pick the lower real solution of this equation ($z=0.09334907$),
since the upper one is an artefact of making the original equation (\ref{2.29}) rational. 
It gives the exact tricritical point of the present GL model 
\begin{equation}
m_{\rm tri}= 0.825765 , \quad \nu_{\rm tri} = 2.935050, \quad \sigma_{\rm tri}= 1.635108 ,
\label{2.31}
\end{equation}
superseding the previous numerical values (\ref{2.25}). 
Undoing the scale transformation, the asymptotic behavior of the tricritical line in ($\gamma,\mu,T$) space for  $\gamma \to 0$ is
then given by 
\begin{equation}
\mu_{\rm tri} = \frac{1}{2} \nu_{\rm tri} \gamma^{1/3} , \quad T_{\rm tri} = T_c \left( 1 - \frac{1}{a} \sigma_{\rm tri}^2 \gamma^{2/3} \right).
\label{2.32}
\end{equation}
An important step in the above derivation is to use $q,R$ from the LO calculation, thereby reducing the NLO problem to one
with 2 parameters only. Since this will be crucial for applications of a similar method beyond
the regime of GL theory, we have verified carefully that it is indeed correct. In the present model one can eliminate all 3 parameters
$A,B,q$ exactly by minimizing the full effective potential. It is necessary to solve a cubic equation on the way, so that the 
steps towards finding the tricritical point are more involved. Nevertheless, we could fully confirm the result
of the simpler calculation presented above so that we are confident that the basic idea is correct.

In the practical application of this method to a HF calculation, the use of the Hesse matrix (\ref{2.27}) has turned out to be problematic. Alternatively
one can first minimize the effective action with respect to $B$. Then one looks for the vanishing of the 2nd derivative with respect to $m$ of
the remaining effective potential, now a function of one variable only. This is the way we shall implement the NLO stability analysis in the HF framework below.
In the GL case, this is perhaps less elegant but again reproduces the above result exactly.

Finally, let us have a look at the values of $R$ and $q$ characterizing the unstable mode. At the tricritical point, we find
\begin{equation}
q = 0.997 \nu, \quad R=0.924
\label{2.33}
\end{equation}
$q/\nu$ and $R$ both approach 1 monotonically for $m\to 0$ or $\nu \to \infty$, the values characteristic for the chiral spiral. The deviation
from these limiting values is remarkably small, in particular for $q/\nu$.

Summarizing what we have learned from the GL model, we have found a way to locate the tricritical point by approaching it from
the perturbative side. As one might have expected, it is necessary to push PT to NLO, but only in the direction of
the LO unstable mode. In the next section, we therefore turn to NLO PT of the fermion spectrum as needed for a 
full HF calculation.

%<<<<<<<<<<<<<<<<<<<<<<<<<<<<<<<<<<<<<<<<<<<<<<<<<<<<<<<<<<<<<<<<<<<<<<<<<<<<<<<<<<<<<<<<<<<< <<<<<<<<<<<<<<<<<<<<<<<<<<<<<
%<<<<<<<<<<<<<<<<<<<<<<<<<<<<<<<<<<<<<<<<<<<<<<<<<<<<<<<<<<<<<<<<<<<<<<<<<<<<<<<<<<<<<<<<<<<<<<<<<<<<<<<<<<<<<<<<<<<<<<<<<<
\section{Perturbative fermion spectrum for Hartree-Fock calculation}
\label{sect3}
%<<<<<<<<<<<<<<<<<<<<<<<<<<<<<<<<<<<<<<<<<<<<<<<<<<<<<<<<<<<<<<<<<<<<<<<<<<<<<<<<<<<<<<<<<<<<<<<<<<<<<<<<<<<<<<<<<<<<<<<<<<
%<<<<<<<<<<<<<<<<<<<<<<<<<<<<<<<<<<<<<<<<<<<<<<<<<<<<<<<<<<<<<<<<<<<<<<<<<<<<<<<<<<<<<<<<<<<<<<<<<<<<<<<<<<<<<<<<<<<<<<<<<<

The perturbative phase boundary can be found by a stability analysis. To this end, it is sufficient to compute the fermion spectrum
for massive Dirac fermions subject to a complex, harmonic perturbation to LO, i.e., to 2nd order in the strength of the inhomogeneous potential.
For the present study of the tricritical point, it is mandatory to extend this calculation to NLO (4th order PT). Naive PT
breaks invariably down near the gaps characteristic for periodic potentials. This problem shows up already in LO, and we shall review the manner in
which it has been solved there.  NLO PT presents additional challenges to be addressed in the present section. 

The unperturbed Hamiltonian is the free, massive Dirac Hamiltonian with the familiar spectrum of positive ($\eta=1$) and negative ($\eta=-1$) energy states,
\begin{eqnarray}
H_0 & = & -i \gamma_5 \partial_x + \gamma^0 m,
\nonumber \\
H_0 |\eta, p \rangle & = & \eta \sqrt{m^2+p^2} |\eta, p \rangle .
\label{3.1}
\end{eqnarray}
With the choice of $\gamma$-matrices 
\begin{equation}
\gamma^0 = \sigma_1, \quad \gamma^1 = i \sigma_2, \quad \gamma_5 = \gamma^0 \gamma^1 = - \sigma_3,
\label{3.2}
\end{equation}
the free spinors read
\begin{equation}
\langle x | \eta, p \rangle  =  \frac{1}{\sqrt{2E}}\left( \begin{array}{c} \sqrt{E-\eta p} \\ \eta \sqrt{E+\eta p}\end{array} \right) e^{ipx},  \quad E=\sqrt{m^2+p^2} .
\label{3.3}
\end{equation}
A perturbation appropriate for the stability analysis of the $\chi$GN model is
\begin{equation}
V=\gamma^0 2 S_1 \cos (2Qx) -i \gamma^1 2P_1 \sin (2Qx).
\label{3.4}
\end{equation}
Matrix elements of $V$ in the unperturbed basis can be evaluated analytically,
\begin{eqnarray}
\langle \eta',p'| V | \eta,p \rangle & = &  \frac{1}{2\sqrt{E E'}} \left( {\cal A}^{(+)}\delta^{(+)} S_1 + 
{\cal A}^{(-)}\delta^{(-)} P_1\right),
\nonumber \\
{\cal A}^{(\pm)} & = & \eta \sqrt{E+\eta p} \sqrt{E'-\eta'p'} \pm \eta' \sqrt{E-\eta p}\sqrt{E'+\eta'p'},
\nonumber \\
\delta^{(\pm)} & = & \delta_{p',p-2Q} \pm\delta_{p',p+2Q}, \quad E=\sqrt{m^2+p^2}, \quad E' = \sqrt{m^2+(p')^2}.
\label{3.5}
\end{eqnarray}
The perturbing potential connects only states with momenta differing by $\pm 2 Q$,  having the same or the opposite $\eta$.
Thus, when dealing with PT, the relevant momenta can be considered as discrete, and the $\delta's$ in (\ref{3.5}) as Kronecker delta's.

%<<<<<<<<<<<<<<<<<<<<<<<<<<<<<<<<<<<<<<<<<<<<<<<<<<<<<<<<<<<<<<<<<<<<<<<<<<<<<<<<<<<<<<<<<<<<<<<<<<<<<<<<<<<<<<<<<<<<<<<<<<
\subsection{Fourth order non-degenerate perturbation theory}
\label{sect3A}
%<<<<<<<<<<<<<<<<<<<<<<<<<<<<<<<<<<<<<<<<<<<<<<<<<<<<<<<<<<<<<<<<<<<<<<<<<<<<<<<<<<<<<<<<<<<<<<<<<<<<<<<<<<<<<<<<<<<<<<<<<<

Sufficiently far away from the gaps in the spectrum, we may use standard non-degenerate PT. Since this is covered in any textbook on quantum mechanics, we can be
very brief. As $V$ acts only between states with momenta differing by $\pm 2Q$, only even powers of $V$ lead back to the same state. We exploit this
fact in order to simplify the formulae in the present subsection. We multiply $V$ by a formal parameter $\lambda$ and consider the energy eigenvalues to
LO ($\lambda^2)$ and NLO ($\lambda^4$).

The simplest derivation of Rayleigh-Schr\"odinger PT starts from Brillouin-Wigner PT for the wave function,
\begin{equation}
|\psi_n \rangle = \left[1 + \lambda G_n^{(0)}(E_n) V + \lambda^2 (G_n^{(0)}(E_n)V)^2  + \lambda^3 (G_n^{(0)}(E_n)V )^3 \right] |\psi_n^{(0)}\rangle
\label{3.6}
\end{equation} 
with
\begin{equation}
G_n^{(0)}(E)= \frac{Q_n}{E - H_0}, \quad Q_n = 1-P_n, \quad P_n = |\psi_n^{(0)}\rangle \langle \psi_n^{(0)}|.
\label{3.7}
\end{equation}
The exact energy eigenvalue $E_n$ has the expansion 
\begin{equation}
E_n - E_n^{(0)}  = \lambda \langle \psi_n^{(0)}| V| \psi_n \rangle =  \lambda^2 E_n^{(2)} + \lambda^4 E_n^{(4)}.
\label{3.8}
\end{equation}
Here, $E_n^{(0)}$ denotes the unperturbed energies, $|\psi_n^{(0)}\rangle$ the unperturbed state vector. The Rayleigh-Schr\"odinger result can then be obtained 
by expanding the Green's functions up to the required order 
\begin{equation}
G_n^{(0)}(E_n) = g_n - \lambda^2 E_n^{(2)} g_n^2  , \quad g_n =G_n^{(0)}(E_n^{(0)}).
\label{3.9}
\end{equation}
 Up to 4-th order, one finds
\begin{eqnarray}
E_n^{(2)} & = & \langle \psi_n^{(0)} | V g_n V |\psi_n^{(0)} \rangle ,
\nonumber \\
E_n^{(4)} & = & \langle \psi_n^{(0)} | V g_n V g_n V g_n V |\psi_n^{(0)} \rangle - E_n^{(2)} \langle \psi_n^{(0)} | V g_n^2 V | \psi_n^{(0)} \rangle.
\label{3.10}
\end{eqnarray}
If one would insert the spectral decomposition of the Green's functions, one would generate a large number of individual contributions (4 LO and 48 NLO terms). We did not evaluate each
contribution separately and sum them up, but used computer algebra (Maple) to generate the result automatically. The method used will be explained in subsection \ref{sect3C}.
Without calculation, it is clear that there are singular terms. Let us label the unperturbed state $|\eta, p\rangle$ by ($\eta,0$) and an excited
state $|\eta', p+2nQ\rangle$ by ($\eta',n$). Then the state $(\eta,0)$ is (nearly) degenerate with $(\eta,\pm1)$ near $p=\mp Q$ and with $(\eta, \pm 2)$ near $p=\mp 2Q$, the 
positions of the lowest two gaps. Accordingly there is a first order pole in $E^{(2)}$ corresponding to the transitions $(\eta,0)\to (\eta,\pm 1) \to (\eta, 0)$. In $E^{(4)}$, 
a similar pole shows up in the four-step processes $(\eta,0) \to (\eta', \pm 1 ) \to (\eta,\pm2)\to (\eta'',\pm 1) \to (\eta,0)$. In addition, there are 2nd order and even 3rd order poles
coming either from the process $(\eta,0)\to (\eta',\pm 1) \to (-\eta,0) \to (\eta'',\pm 1)\to (\eta,0)$, or from the last term in (\ref{3.10}) with the squared Green's function $g_n^2$.
The resulting spectrum will be illustrated below, but evidently cannot be trusted in the vicinity of
the gaps at $p=\pm Q,\pm 2Q$. If we would proceed to even higher order, additional singularities would arise as further gaps open up at $p = \pm nQ, n>2$.

This problem is of course well known and has already been met in the case of a standard stability analysis. The cure is also known --- ADPT.
Actually, in the LO case there is a cheap way out: as noted in \cite{L18}, for the purpose of finding the phase boundary, non-degenerate LO 
PT can be used on condition that one treats the singularity at the gap by the principal value prescription. We shall explain intuitively why this works in the next section.
When trying to find the tricritical point using a NLO stability analysis, we do not expect any such shortcut as there is no principal value prescription for 2nd or 3rd order poles.
This forces us to consider seriously higher order ADPT. Non-degenerate PT to NLO is still useful sufficiently far away from the gaps where it has actually been
used in the final computations.

%<<<<<<<<<<<<<<<<<<<<<<<<<<<<<<<<<<<<<<<<<<<<<<<<<<<<<<<<<<<<<<<<<<<<<<<<<<<<<<<<<<<<<<<<<<<<<<<<<<<<<<<<<<<<<<<<<<<<<<<<<<
\subsection{Fourth order almost degenerate perturbation theory}
\label{sect3B}
%<<<<<<<<<<<<<<<<<<<<<<<<<<<<<<<<<<<<<<<<<<<<<<<<<<<<<<<<<<<<<<<<<<<<<<<<<<<<<<<<<<<<<<<<<<<<<<<<<<<<<<<<<<<<<<<<<<<<<<<<<<

We now turn to higher order ADPT. Beyond LO, various systematic methods are available, less widely known than 
lowest order ADPT. We choose a convenient scheme due to Lindgren \cite{L27}, originally developed for many body calculations in chemistry and nuclear physics. 
It has the advantage of being based on Rayleigh-Schr\"odinger PT and can be described as follows: Those states which mix strongly with a given state are said to belong to a subspace of 
Hilbert space called $P$-space (or model space). All other states define the $Q$-space. The number of states in $P$-space is not limited a priori. 
In the present application, $P$ space is always 2-dimensional since only one pair of states is degenerate at the gap positions. Thus near the lower gap,
$P$ space will be chosen as the pair of states $|\eta, p\rangle,|\eta,p\pm 2Q\rangle$ near $p=\mp Q$, respectively. Near the upper gap, it will contain the states $|\eta, p\rangle, |\eta,p\pm 4Q\rangle$
near $p=\mp 2Q$. Then the formalism allows one to construct an effective Hamiltonian in $P$-space whose eigenvalues approach the exact energies of 2 states in the 
full Hilbert space as one increases the order of perturbation. Since the effective Hamiltonian can be diagonalized exactly, the divergences of naive
PT are avoided.   

We only state the necessary steps up to 4th order PT, referring to the original work \cite{L27} for the derivation and background. Let us label unperturbed states in
$P$-space by Latin letters and states in $Q$-space by Greek letters, where the labels include the momentum $p$ and the sign of the energy $\eta$. 
Then the effective Hamiltonian can be written down concisely as
\begin{equation}
H_{\rm eff} = P H_0 P + \lambda P V \Omega
\label{3.11}
\end{equation}
with 
\begin{eqnarray}
\Omega & = & P + \lambda \Omega^{(1)} + \lambda^2 \Omega^{(2)} + \lambda^3 \Omega^{(3)},
\nonumber \\
\Omega^{(1)}_{\alpha j} & = & \frac{V_{\alpha j}}{e_j - e_{\alpha}},
\nonumber \\
\Omega^{(2)}_{\alpha j} & = & \frac{\sum_{\beta} V_{\alpha \beta}\Omega^{(1)}_{\beta j} - \sum_k \Omega^{(1)}_{\alpha k} V_{kj}}{e_j - e_{\alpha}},
\nonumber \\
\Omega^{(3)}_{\alpha j} & = & \frac{\sum_{\beta} V_{\alpha \beta} \Omega^{(2)}_{\beta j} - \sum_{k,\beta} \Omega_{\alpha k}^{(1)} V_{k \beta} \Omega_{\beta j}^{(1)} -  \sum_k \Omega^{(2)}_{\alpha k} V_{k j}}{e_j-e_{\alpha}}.
\label{3.12}
\end{eqnarray}
$P$ is the projector onto $P$-space. The $\Omega^{(n)}$ are kind of wave operators leading from $P$ to $Q$-space $(\Omega_{\alpha j})$, defined iteratively. $V$ acts within $P$ space ($V_{kj}$), within
$Q$-space ($V_{\alpha \beta}$) and between $P$ and $Q$ space ($V_{\alpha j},V_{k\beta})$. Inserting the lower $\Omega$'s successively into the hierarchy (\ref{3.12}), we arrive at the effective Hamiltonian
\begin{eqnarray}
(H_{\rm eff})_{ij} & = & \delta_{ij} e_i+ \lambda V_{ij} + \lambda^2 \sum_{\alpha}  \frac{V_{i\alpha} V_{\alpha j}}{e_j-e_{\alpha}}
\nonumber \\
& + & \lambda^3  \sum_{\alpha \beta} \lambda^3 \frac{V_{i \alpha} V_{\alpha \beta} V_{\beta j}}{(e_j-e_{\alpha})(e_j-e_{\beta})} - \lambda^3 \sum_{\alpha k} \frac{V_{i \alpha} V_{\alpha k} V_{k j}}{(e_j-e_{\alpha})(e_k-e_{\alpha})} 
\nonumber \\
& + & \lambda^4 \sum_{\alpha \beta \gamma} \frac{V_{i \alpha} V_{\alpha \beta} V_{\beta \gamma} V_{\gamma j}}{(e_j-e_{\alpha})  (e_j-e_{\beta})  (e_j-e_{\gamma}) }
\nonumber \\
& - & \lambda^4  \sum_{\alpha \beta k}  \frac{V_{i \alpha} V_{\alpha \beta} V_{\beta k} V_{k j}}{(e_j-e_{\alpha})  (e_j-e_{\beta})  (e_k-e_{\beta}) }
\nonumber \\
& - & \lambda^4 \sum_{\alpha \beta k} \frac{V_{i \alpha} V_{\alpha k} V_{k \beta} V_{\beta j}}{(e_j-e_{\alpha})  (e_k-e_{\alpha})  (e_j-e_{\beta}) }
\nonumber \\
& - & \lambda^4 \sum_{\alpha \beta k} \frac{V_{i \alpha} V_{\alpha \beta} V_{\beta k} V_{k j}}{(e_j-e_{\alpha})  (e_k-e_{\alpha})  (e_k-e_{\beta}) }
\nonumber \\
& + & \lambda^4 \sum_{\alpha k \ell} \frac{V_{i \alpha} V_{\alpha \ell} V_{\ell k} V_{k j}}{(e_j-e_{\alpha})  (e_k-e_{\alpha})  (e_{\ell}-e_{\alpha}) }.
\label{3.13}
\end{eqnarray}
There are no vanishing energy denominators anymore (terms with two Latin labels in the same energy denominator). Each pair of strongly mixed states is treated non-perturbatively by diagonalizing
the  2$\times$2 matrix $H_{\rm eff}$. 

Before turning to the actual calculation, it is worth pointing out some simplifications. For the lower gap region where the potential can connect the two states in $P$ space, only even terms survive
($\lambda^2, \lambda^4$) in diagonal matrix elements, odd terms ($\lambda,\lambda^3$) in off-diagonal elements. This is a direct consequence of the fact that the potential 
acts as ladder operator in momentum space.  
For the upper gap, all terms with $V_{ij}$ can be dropped because the potential does not connect directly the two states in $P$-space.  Here, all matrix elements of $H_{\rm eff}$ are even in $\lambda$.

%<<<<<<<<<<<<<<<<<<<<<<<<<<<<<<<<<<<<<<<<<<<<<<<<<<<<<<<<<<<<<<<<<<<<<<<<<<<<<<<<<<<<<<<<<<<<<<<<<<<<<<<<<<<<<<<<<<<<<<<<<<
\subsection{Calculational method and illustration of the spectrum}
\label{sect3C}
%<<<<<<<<<<<<<<<<<<<<<<<<<<<<<<<<<<<<<<<<<<<<<<<<<<<<<<<<<<<<<<<<<<<<<<<<<<<<<<<<<<<<<<<<<<<<<<<<<<<<<<<<<<<<<<<<<<<<<<<<<<
%###########################################################################################################################
\begin{figure}
\begin{center}
\epsfig{file=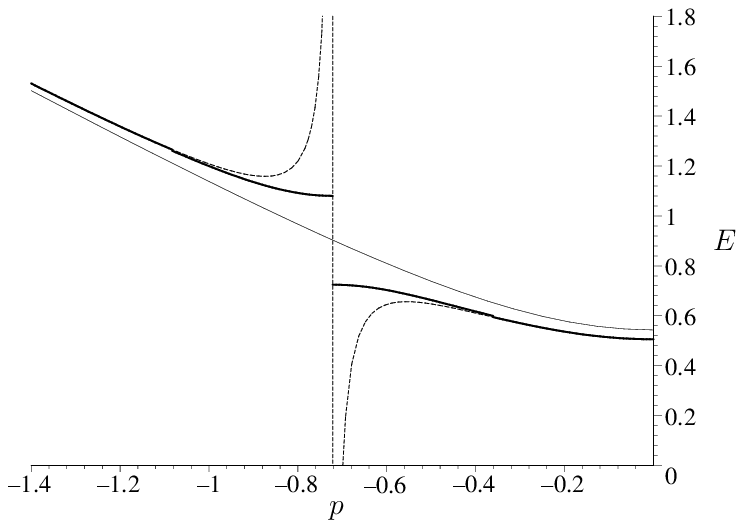,height=6cm,width=8cm,angle=0}
\caption{Typical fermion spectrum in the region of the lower gap at $p=-Q$, $E>0$. Thin solid line: Unperturbed energy, dashed line: naive LO PT,
fat solid line: ADPT using Lindgren's method to LO.} 
\label{fig2}
\end{center}
\end{figure}
%####################################################################################################\#######################

As pointed out above, the LO stability analysis requires only 2nd order naive PT and the principal value prescription for divergent integrals. This is so simple 
that one can write down the resulting spectrum in closed analytical form \cite{L18}. By contrast, inspection of the 4th order ADPT formalism is quite sobering at first sight.
Due to the presence of positive and negative energies and the complexity of the formalism, one finds a huge number of individual
contributions to the effective Hamiltonian. For both gaps together one needs a total of 18 second order terms, 16 third order terms and
242 fourth order terms. Here, an $n$-th order term consists of a product of $n$ matrix elements of $V$ as given in Eq.~(\ref{3.5}), divided by a product of $n-1$ energy 
denominators. Clearly, one cannot expect any presentable closed form result, although the calculation is fully analytical. We have found
a simple way of generating all these many terms automatically with Maple. Starting from the state $|\eta,p\rangle$ and restricting 
ourselves to $p<0$ (the spectrum is symmetric in $p$), this state is connected to 6 other states at most in a NLO ADPT calculation, namely $|\eta', p+2nQ\rangle, n=-2,-1,1,2,3,4$.
Taking into account positive and negative energy states, this implies that a 4th order ADPT calculation of the base state $|\eta,p\rangle$ always stays within a 14-dimensional
subspace of Hilbert space. We have therefore set up the matrix $V$ in this subspace, using the matrix elements given above, in algebraic form. Likewise, the projection
operators and free Green's functions entering the construction of $H_{\rm eff}$ can be set up analytically as matrices. We then literally perform the steps in Lindgren's iterative method (\ref{3.12})
with Maple, generating the effective Hamiltonian automatically. Since only matrix multiplications are required, there is no problem in
getting the result algebraically. Likewise, diagonalization of $H_{\rm eff}$ can be done exactly since this is only a 2$\times$2 matrix. It would be impossible to print the resulting formulas here,
but they can be used within Maple for exact numerical computations by just plugging in numbers when needed and performing one-dimensional integrations numerically. There is no loss
of accuracy as compared to a LO calculation. In a similar manner we have generated all the terms in non-degenerate 4th order PT with little effort and greatly reduced risk of mistakes.

We now illustrate results for the spectrum in a few cases. Consider first the LO calculation needed to find the phase boundary between homogeneous and inhomogeneous phases.
In this case, only the terms of order $\lambda,\lambda^2$ are kept in Eq.~(\ref{3.13}).  
In Fig.~\ref{fig2}, we show the spectrum for positive energy states in the vicinity of the first gap, for $p<0$. The parameters chosen are in the range used in 
our calculations, with $Q=0.72$. Naive PT exhibits the expected singularity at the gap, whereas ADPT produces a finite gap non-perturbatively.
In spite of the drastic failure of naive PT seen in this figure, it is worth noting that previous LO stability analyses were carried out using the spectrum from naive PT
together with the principal value prescription. The reason why this is allowed is the following. A LO stability analysis amounts to doing 2nd order PT, but only the limit $\lambda\to 0$
is actually used. The example shown in Fig.~\ref{fig2} corresponds to the choice $\lambda=0.1$. What happens if we start to decrease $\lambda$? The region where naive PT 
differs significantly from ADPT shrinks and the gap tends to 0. 
%###########################################################################################################################
\begin{figure}
\begin{center}
\epsfig{file=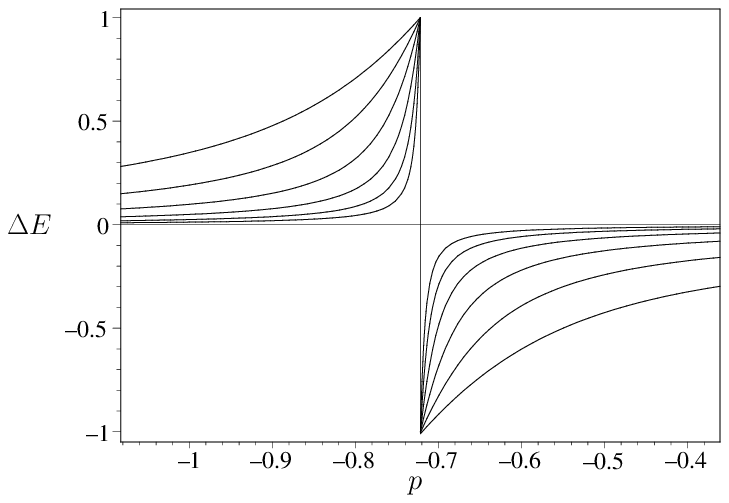,height=6cm,width=8cm,angle=0}
\caption{Correction to unperturbed spectrum near lower gap in LO ADPT, normalized to $\pm 1$ at $p=-Q$. The different curves show the effect of decreasing 
$\lambda$, see main text. In the limit $\lambda\to 0$, the region where the gap is being felt shrinks to 0.} 
\label{fig3}
\end{center}
\end{figure}
%####################################################################################################\#######################
This is illustrated in Fig.~\ref{fig3} for the values $\lambda=0.1/2^n, n=0...5$. The curves show the correction to the unperturbed spectrum and
are strictly antisymmetric with respect to the gap position $-Q$, as one can verify by inspection of the underlying formulas. They have been normalized to $\pm 1$ at the gap ($p=-Q$ )
so as to show more clearly how the width of the gap region gets contracted as $\lambda\to 0$. In the limit $\lambda \to 0$, multiplying such a function by a smooth function and 
integrating over it is nothing but a principal value integral. There is thus no loss of accuracy using
this trick. Since we now have at our disposition the possibility to do a LO ADPT computation explicitly for finite $\lambda$, we have checked this intuitive
reasoning numerically to high accuracy. 
%###########################################################################################################################
\begin{figure}
\begin{center}
\epsfig{file=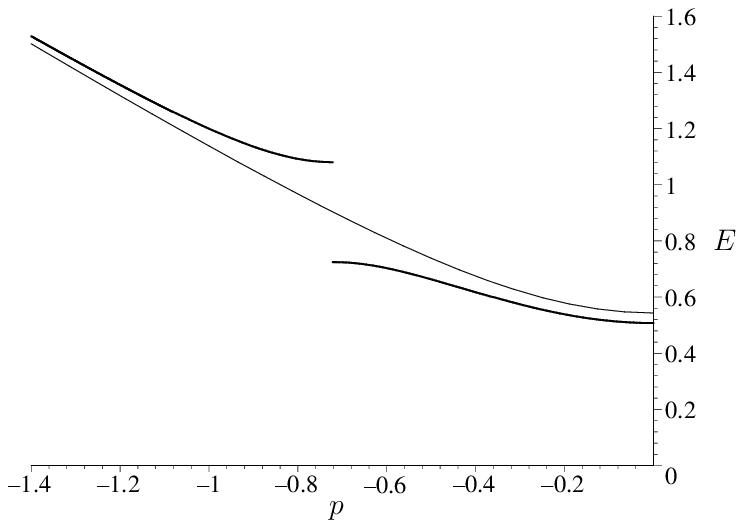,height=6cm,width=8cm,angle=0}
\caption{Example of spectrum obtained with Lindgren's method to LO and NLO. The thin line is the unperturbed energy. The positive energy region around
the lower gap is shown for typical values of the parameters.} 
\label{fig4}
\end{center}
\end{figure}
%####################################################################################################\#######################

Let us now turn to an example of the full NLO spectrum in the regions where naive PT fails. In Figs. \ref{fig4} and \ref{fig5} we show the spectrum for $E>0$ in the vicinity of 
the lower gap ($p=-Q$) and the upper gap ($p=-2Q$). We do not show the corresponding LO calculation which does not even produce the upper gap.
Near the lower gap, the corrections are actually rather small.

%###########################################################################################################################
\begin{figure}
\begin{center}
\epsfig{file=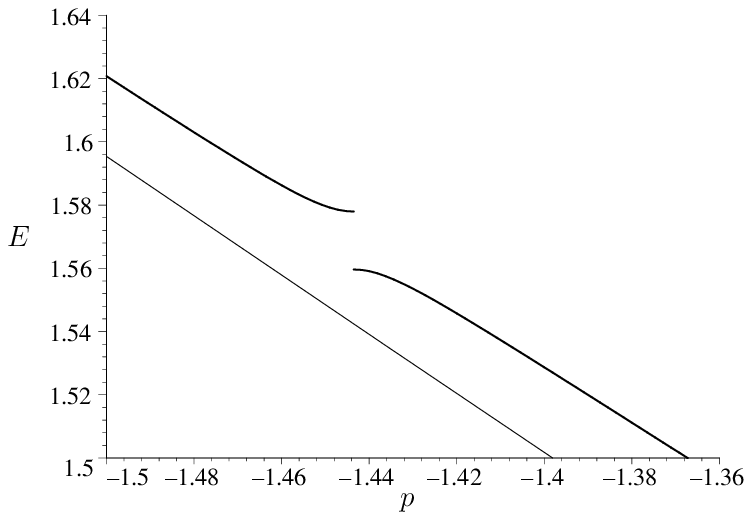,height=6cm,width=8cm,angle=0}
\caption{Like Fig.~\ref{fig4}, but for the region around the upper gap.} 
\label{fig5}
\end{center}
\end{figure}
%####################################################################################################\#######################
%###########################################################################################################################
\begin{figure}
\begin{center}
\epsfig{file=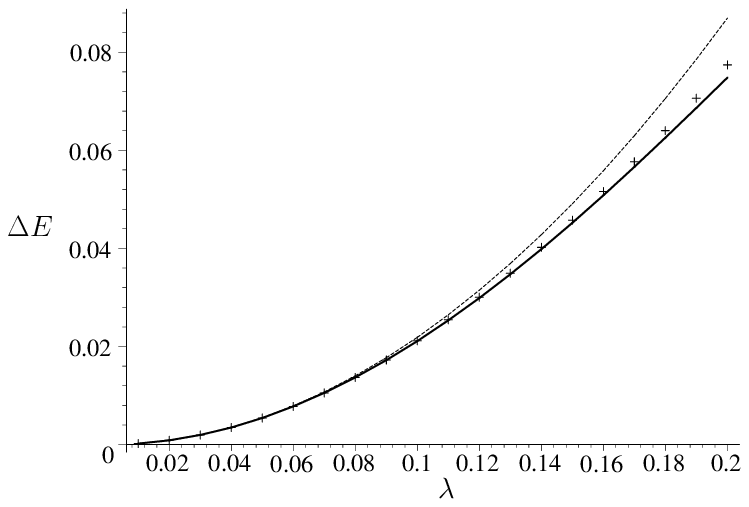,height=6cm,width=8cm,angle=0}
\caption{Test of Lindgren's analytical approach to LO (dashed curve) and NLO (solid curve) against exact numerical eigenvalues (crosses) of the Hamiltonian, 
truncated to a a 14 dimensional subspace of Hilbert space. One particular eigenstate is chosen, and all parameters including $p$ are fixed except for the
strength parameter $\lambda$.} 
\label{fig6}
\end{center}
\end{figure}
%####################################################################################################\#######################
%###########################################################################################################################
\begin{figure}
\begin{center}
\epsfig{file=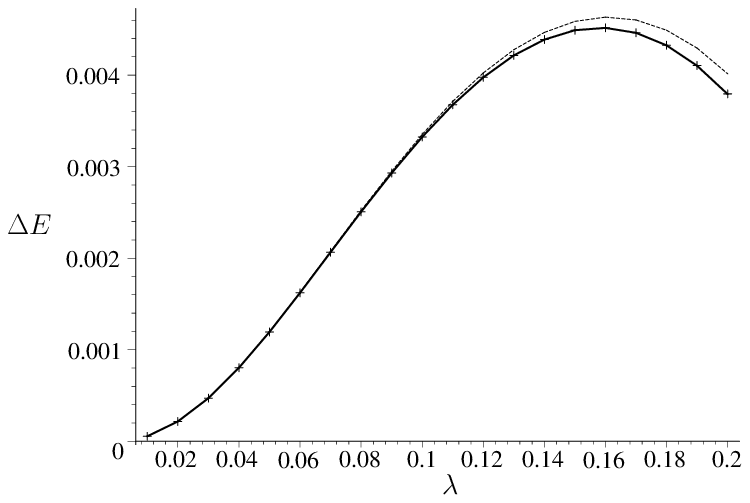,height=6cm,width=8cm,angle=0}
\caption{Same as Fig.~\ref{fig6}, but for another eigenstate.} 
\label{fig7}
\end{center}
\end{figure}
%####################################################################################################\#######################
Finally we should like to remark that our method of generating all perturbative contributions automatically with Maple can also be used for an additional test.
To check whether the NLO calculation really improves the LO calculation, one would ideally need the exact eigenvalues of the 
Hamiltonian. What we have done instead is to diagonalize numerically the 14-dimensional Hamiltonian submatrix used in the computer algebra calculations.
The perturbative expressions should converge to two of the exact eigenvalues of this matrix in the limit  $\lambda \to 0$. These are not the same as the
eigenvalues in the full Hilbert space, but provide the means to test the perturbative scheme in a simple manner. In the region of parameters needed here,
the results are very good indeed. In Figs.~\ref{fig6} and \ref{fig7}, we show two examples, plotting the difference of exact eigenvalues and unperturbed energies 
as a function of $\lambda$ in comparison with perturbative results. 
The crosses are from the numerical diagonalization, the dashed  lines LO ADPT and the solid lines NLO ADPT. The important point here is that the NLO 
calculation really improves the LO result. The differences may seem small, but we should not forget that the determination of the tricritical point 
depends critically just on this difference.
Together with many similar plots, Figs.~\ref{fig6} and \ref{fig7} give us confidence that Lindgren's method works well in the regime we are intersted in.

%<<<<<<<<<<<<<<<<<<<<<<<<<<<<<<<<<<<<<<<<<<<<<<<<<<<<<<<<<<<<<<<<<<<<<<<<<<<<<<<<<<<<<<<<<<<< <<<<<<<<<<<<<<<<<<<<<<<<<<<<<
%<<<<<<<<<<<<<<<<<<<<<<<<<<<<<<<<<<<<<<<<<<<<<<<<<<<<<<<<<<<<<<<<<<<<<<<<<<<<<<<<<<<<<<<<<<<<<<<<<<<<<<<<<<<<<<<<<<<<<<<<<<
\section{Locating the tricritical point}
\label{sect4}
%<<<<<<<<<<<<<<<<<<<<<<<<<<<<<<<<<<<<<<<<<<<<<<<<<<<<<<<<<<<<<<<<<<<<<<<<<<<<<<<<<<<<<<<<<<<<<<<<<<<<<<<<<<<<<<<<<<<<<<<<<<
%<<<<<<<<<<<<<<<<<<<<<<<<<<<<<<<<<<<<<<<<<<<<<<<<<<<<<<<<<<<<<<<<<<<<<<<<<<<<<<<<<<<<<<<<<<<<<<<<<<<<<<<<<<<<<<<<<<<<<<<<<<

We now come to the central part of this work. Given that we are able to compute the spectrum of the HF Hamiltonian in LO and NLO reliably,
how do we find the tricritical point? We shall use the experience with the GL model as guideline. We first have to determine the grand canonical potential
perturbatively as a function of ($\mu,\gamma,T$). In HF approximation, it is given by a single particle contribution plus a double counting correction,
\begin{equation}
\Psi = - \frac{2}{\beta} \int_0^{\Lambda/2} \frac{dp}{2\pi} \ln \left[ (1+e^{-\beta(E_{1,p}-\mu)})(1+e^{-\beta(E_{-1,p}-\mu)})\right] + \frac{(m-m_b)^2+2(S_1^2+P_1^2)}{2N g^2}.
\label{4.1}
\end{equation}
$E_{\pm 1,p}$ are positive and negative single particle energies belonging to the potential (\ref{3.4}). The bare coupling constant will be eliminated
with the help of the vacuum gap equation
\begin{equation}
\frac{\pi}{Ng^2} = \gamma+ \ln \Lambda .
\label{4.2}
\end{equation}
To isolate the UV divergence of the integral, we separate the vacuum from the matter contributions,
\begin{eqnarray}
\Psi & = & \Psi_{\rm vac} + \Psi_{\rm matt} + \frac{(m-m_b)^2+2(S_1^2+P_1^2)}{2N g^2},
\nonumber \\
\Psi_{\rm vac} & = & 2 \int_0^{\Lambda/2} \frac{dp}{2\pi} E_{-1,p},
\nonumber \\
\Psi_{\rm matt} & = & - \frac{2}{\beta} \int_0^{\Lambda/2} \frac{dp}{2\pi} \ln \left[ (1+e^{-\beta(E_{1,p}-\mu)})(1+e^{\beta(E_{-1,p}-\mu)})\right] .
\label{4.3}
\end{eqnarray}
The single particle energies are expanded into a perturbation series up to NLO, 
\begin{equation}
E_{\eta,p} = E_{\eta,p}^{(0)}  + \epsilon E_{\eta,p}^{(1)} + \epsilon^2  E_{\eta,p}^{(2)}. 
\label{4.4}
\end{equation}
In the double counting correction, $S_1^2+P_1^2$ has to be treated as being of order $\epsilon$. 
One then expands the grand canonical potential into a Taylor series in $\epsilon$. 
The 0-th order term reproduces what one would write down for the homogeneous phase diagram. The linear terms are ($E=\sqrt{m^2+p^2}$)
\begin{eqnarray}
\Psi_{\rm vac}^{(1)} & = & 2 \int_0^{\Lambda/2} \frac{dp}{2\pi} E_{-1,p}^{(1)},
\nonumber \\
\Psi_{\rm matt}^{(1)} & = &  \int_0^{\infty} \frac{dp}{\pi}  \left( \frac{E_{1,p}^{(1)}}{e^{\beta(E-\mu)}+1} - \frac{E_{-1,p}^{(1)}}{e^{\beta(E+\mu)}+1} \right) .
\label{4.5}
\end{eqnarray}
This part is used to find the perturbative phase boundary in a standard LO stability analysis \cite{L18}.
The vacuum term has to be renormalized with the help of the O($\epsilon$) double counting correction,  
\begin{equation}
\Psi_{\rm vac}^{(1)} + \frac{S_1^2+P_1^2}{Ng^2}    =  \frac{S_1^2+P_1^2}{\pi} [\gamma+ \ln (2 K)] + 2 \int_0^{K} \frac{dp}{2\pi} E_{-1,p}^{(1)}.
\label{4.6} 
\end{equation}
Here, $K$ is a momentum chosen such that the integrand can be approximated by the leading asymptotic term ($\sim 1/p$) for momenta $p\ge K$.
The cutoff $\Lambda$ and the bare coupling constant have disappeared owing to the gap equation.
In NLO, we obtain the novel terms crucial for the search of the tricritical point,
\begin{eqnarray}
\Psi_{\rm vac}^{(2)} & = & 2 \int_0^{\infty} \frac{dp}{2\pi} E_{-1,p}^{(2)},
\nonumber \\
\Psi_{\rm matt}^{(2)} & = &  \int_0^{\infty} \frac{dp}{\pi} \left( \frac{E_{1,p}^{(2)} - \beta (E_{1,p}^{(1)})^2/2}{e^{\beta(E-\mu)}+1} - \frac{E_{-1,p}^{(2)} + \beta (E_{-1,p}^{(1)})^2/2}{e^{\beta(E+\mu)}+1} \right) 
\nonumber \\
& & + \beta \int_0^{\infty} \frac{dp}{2\pi} \left[ \left( \frac{E_{1,p}^{(1)}}{e^{\beta(E-\mu)}+1}\right)^2 +  \left( \frac{E_{-1,p}^{(1)}}{e^{\beta(E+\mu)}+1}\right)^2\right].                
\label{4.7}
\end{eqnarray}
Since all integrands are analytically known, free of singularities and the integrals are UV convergent, these expressions can readily be computed with Maple 
with the desired accuracy.

We now proceed as follows. In a first step, we have to determine the perturbative sheet as in Ref.~\cite{L18}, using naive LO PT with the principal value prescription. 
These computations have already been done in \cite{L18} but had to be repeated because the results for $R=S_1/P_1$ and $Q$ characterizing the unstable mode 
had not been stored at that time. This gives the critical curves in ($\mu,T$) plane for a number of $\gamma$ values. 
In the next step, we choose a point near the (low $\mu$) end of such a critical curve and evaluate the full grand canonical potential including 0-th order term, 
LO and NLO corrections. The parameters ($\mu,T,R,Q$) are taken over from the LO calculation of the critical line. The only undetermined parameters in the potential are $P_1$ and $m$.
Denote by $m_0$ the fermion mass on the perturbative sheet, i.e., the fermion mass in the homogeneous phase. Consider three neighboring values for $m$,
$m=m_0-\Delta m, m_0, m_0+\Delta m$ with $\Delta m \ll m_0$. For each of these three mass values, perform the calculation of $\Psi$ for a range of $P_1$ values and
find the minimum,  
\begin{equation}
\widetilde{\Psi}(m) := \min_{P_1} \Psi(m,P_1).
\label{4.8}
\end{equation}
Then the discretization of the 2nd derivative of the effective potential with respect to $m$ at the point considered is
\begin{equation}
\left. \partial_m^2 \widetilde{\Psi}(m)\right|_{m_0} = \frac{\widetilde{\Psi}(m_0 +\Delta m) - 2 \widetilde{\Psi}(m_0) + \widetilde{\Psi}(m_0-\Delta m)}{\Delta m^2}.
\label{4.9}
\end{equation}
This is the quantity that must vanish at the tricritical point. In practice, it is sufficient to compute (\ref{4.8}) for several equidistant points near the end of the perturbative phase boundary and
determine the zero crossing by interpolation. 

We have done the calculation for all values of $\gamma$ listed in Table~\ref{tab1}. We integrate only over the negative $p$-axis (the integrands are symmetric). Naive 4th order PT has been used 
for $p<-5/2 Q$ and $-1/2 Q<p<0$, away from the gaps. In the interval $-3Q/2<p<-Q/2$ containing the lower gap, we use Lindgren's method with the states $|\eta,p\rangle,|\eta, p+2Q\rangle$
in $P$-space.  In the interval $-5/2Q<p<-3/2Q$ containing the upper gap, the model space consists of $|\eta,p\rangle, |\eta,p+4Q\rangle$. Minimization with respect to $P_1$ presented no
difficulty and the zero of the 2nd derivative (\ref{4.9}) could easily be found in all cases. 
\begin{center}
\begin{table}
\begin{tabular}{|c|c|c|c|c|}
\hline
 $\gamma$ &  $\mu$ & $T$  \\
\hline
0.0125 & .321 & .543 \\
0.025 & .393 & .528 \\
0.05 & .509 & .501 \\
0.1 & .629 & .460 \\
0.2 & .757 & .393 \\
0.3 & .826 & .339 \\
0.4 & .869 & .293 \\
0.5 & .896 & .253 \\
0.6 & .914 & .219 \\
0.7 & .925 & .189 \\
0.8 & .932 & .163 \\
0.9 & .938 & .142 \\
1.0 & .939 & .121 \\
1.1 & .941 & .104 \\
1.2 & .941 & .089 \\
\hline
\end{tabular}
\caption{Numerical values for tricritical points determined in this work}
\label{tab1}
\end{table}
\end{center}
Fig.~\ref{fig8} shows all perturbative phase boundaries and the tricritical line thus obtained. The calculated tricritical points are the end points of the phase boundaries.
They have been connected by straight line segments to guide the eye. The numerical values of the tricritical points are listed in Table~\ref{tab1}, since this is the main result of the present
work. It is interesting to compare the tricritical points obtained with the present ``top down"  NLO stability analysis to the ``bottom up" results obtained in Ref.~\cite{L18}
by a numerical HF calculation and some extrapolation. The result is shown in Figs.~\ref{fig9} and \ref{fig10}. The agreement is surprisingly good, given the fact that the two calculations are completely 
independent. The points from the numerical HF calculation show small fluctuations around the smooth, perturbative curve. This supports the 
claim that it is indeed possible to find the tricritical curves by an exact perturbative calculation, taking into account some non-perturbative aspects related to the gaps. 
Finally, in Fig.~\ref{fig11} we summarize everything known about the phase diagram of the massive $\chi$GN model to date. We have added to the new data from Fig.~\ref{fig8} the
asymptotic curve and the first order sheet from Ref.~\cite{L18}. The $T=0$ line is the baryon mass taken from Ref.~\cite{L28}, and in the vicinity
of the point ($\gamma=0,\mu=0,T=T_c$) we have also included the GL prediction of the present work (the short line segment near $T=T_c,\mu=\gamma=0$). 
The lowest order GL approach is only valid at much smaller values of $\gamma$ then used here ($\gamma \le 0.001$, see \cite{L25}). If one applies it nevertheless to the 
first entry of Table~\ref{tab1} at $\gamma=0.0125$, one finds only qualitative agreement ($\mu=0.341, T=0.553$). 
In any case, Fig.~(\ref{fig8}) underlines that a comprehensive and consistent picture of the phase structure has now been reached. 
%###########################################################################################################################
\begin{figure}
\begin{center}
\epsfig{file=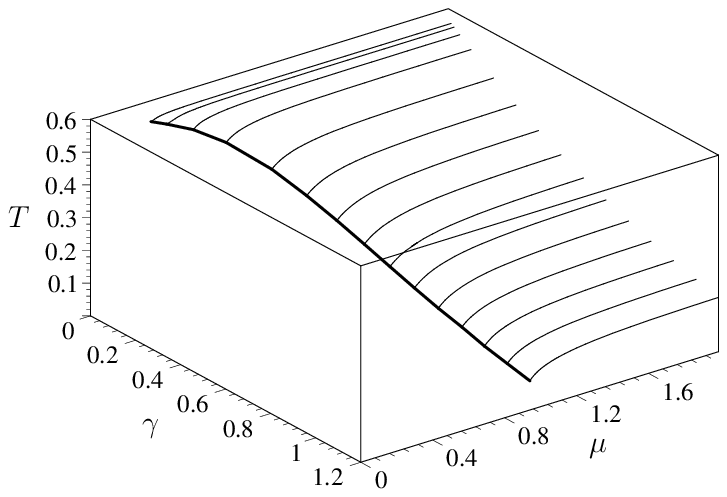,height=6cm,width=8cm,angle=0}
\caption{Result for perturbative sheet and tricritical line obtained in this work using LO and NLO stability analysis, respectively.} 
\label{fig8}
\end{center}
\end{figure}
%####################################################################################################\#######################
%###########################################################################################################################
\begin{figure}
\begin{center}
\epsfig{file=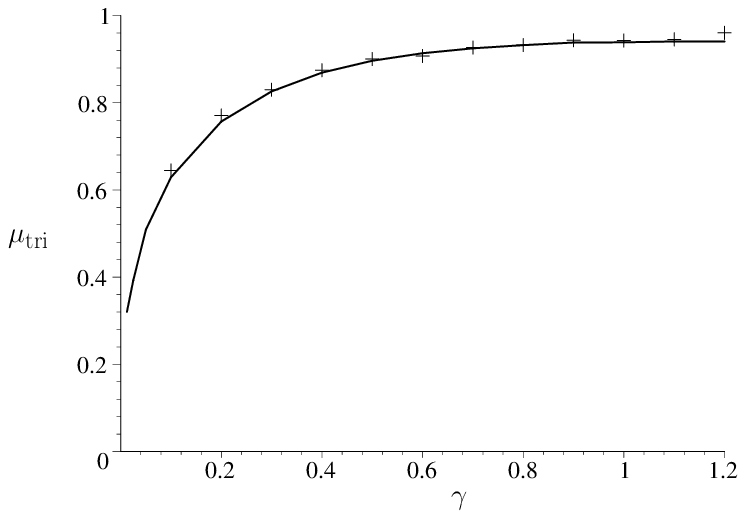,height=6cm,width=8cm,angle=0}
\caption{Comparison of tricritical points obtained from numerical HF calculation (crosses, Ref.~\cite{L18}) and analytical NLO perturbative calculation (solid line),
in the ($\gamma,\mu$) plane.} 
\label{fig9}
\end{center}
\end{figure}
%####################################################################################################\#######################
%###########################################################################################################################
\begin{figure}
\begin{center}
\epsfig{file=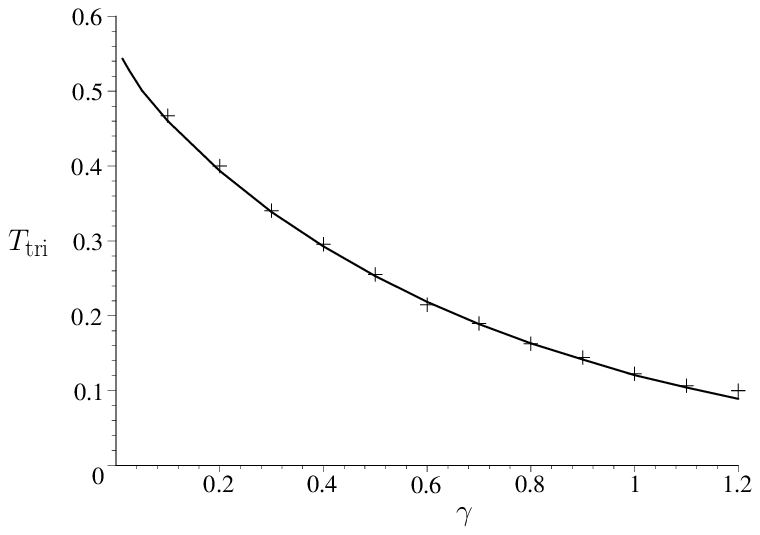,height=6cm,width=8cm,angle=0}
\caption{Like Fig.~\ref{fig9} but ($\gamma,T$) plane shown.} 
\label{fig10}
\end{center}
\end{figure}
%####################################################################################################\#######################
%###########################################################################################################################
\begin{figure}
\begin{center}
\epsfig{file=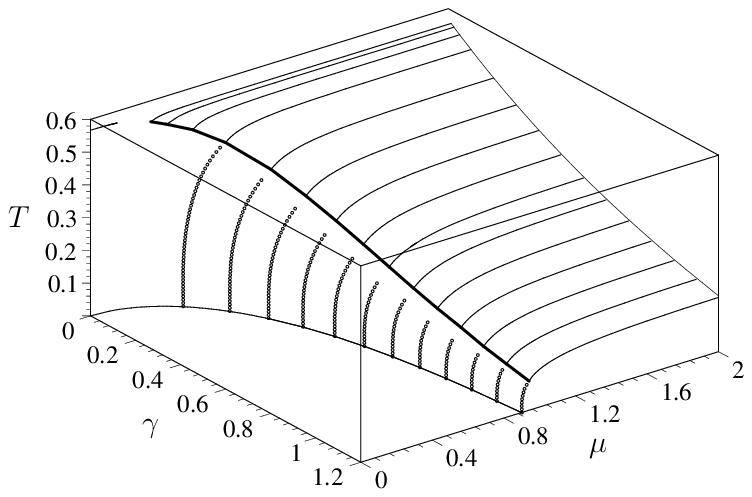,height=10cm,width=13cm,angle=0}
\caption{Summary of everything known about the phase diagram of the massive $\chi$GN model to date, see main text.} 
\label{fig11}
\end{center}
\end{figure}
%####################################################################################################\#######################

%<<<<<<<<<<<<<<<<<<<<<<<<<<<<<<<<<<<<<<<<<<<<<<<<<<<<<<<<<<<<<<<<<<<<<<<<<<<<<<<<<<<<<<<<<<<< <<<<<<<<<<<<<<<<<<<<<<<<<<<<<
%<<<<<<<<<<<<<<<<<<<<<<<<<<<<<<<<<<<<<<<<<<<<<<<<<<<<<<<<<<<<<<<<<<<<<<<<<<<<<<<<<<<<<<<<<<<<<<<<<<<<<<<<<<<<<<<<<<<<<<<<<<
\section{Summary and conclusions}
\label{sect5}
%<<<<<<<<<<<<<<<<<<<<<<<<<<<<<<<<<<<<<<<<<<<<<<<<<<<<<<<<<<<<<<<<<<<<<<<<<<<<<<<<<<<<<<<<<<<<<<<<<<<<<<<<<<<<<<<<<<<<<<<<<<
%<<<<<<<<<<<<<<<<<<<<<<<<<<<<<<<<<<<<<<<<<<<<<<<<<<<<<<<<<<<<<<<<<<<<<<<<<<<<<<<<<<<<<<<<<<<<<<<<<<<<<<<<<<<<<<<<<<<<<<<<<<
Given a phase boundary between a homogeneous and an inhomogeneous phase, how can one find a tricritical point separating first from second order 
transitions? This is the main question addressed in the present work. The specific example which we have studied is the massive $\chi$GN model,
but the basic idea should be applicable to mean field theories in higher dimensions as well. The 2nd order phase boundary can be found in a 
straightforward manner using a stability analysis. This standard tool is based on LO PT in a spatially periodic (``harmonic")  perturbing potential.
It is easy to implement and gives exact results. PT leads to divergencies at the position of a gap, unavoidable for periodic 
perturbations, but these can be handled with well established ADPT methods. The first order phase boundary requires full, numerical HF
calculations and a careful search for the points where two different solutions are degenerate.  So far, the tricritical point could only be determined
by pushing the full HF calculation towards the endpoint of the first order line. This is a lengthy and difficult endeavor, since a weak 1st order transition 
is hard to distinguish from a 2nd order transition numerically. Some ingenuity and a certain amount of extrapolation is needed, so that 
the question about a more efficient location of the tricritical point arises naturally. In the present work, we advocate approaching the tricritical 
point from the perturbative, 2nd order side in a way which is independent of the HF calculation and potentially exact. This requires to
extend the standard LO stability analysis to NLO PT. Due to the periodic perturbation, pushing PT to higher order will
invariably give rise to new divergencies due to vanishing energy denominators. Fortunately, this problem has been solved long time ago in the
context of perturbative many body calculations. We have used a systematic scheme due to Lindgren which is well suited to the present
problem and can be implemented in Maple without much pain. Guided by an exactly solvable warm-up problem, the GL approach to the
$\chi$GN model, we found that it is sufficient to push PT to NLO in the direction dictated by the unstable mode
of the stability analysis. In this way, it was possible to go all the way to locate the tricritical point precisely. In the present case,
the difference between the results from HF and ADPT are rather small, but fluctuations and small systematic errors of the previously determined tricritical points have been eliminated.
The method should be immediately applicable to the $\chi$GN model with isospin, where the tricritical lines are still completely undetermined.
This will be the subject of a forthcoming paper.


\begin{thebibliography}{99}
\bibitem{L1}
D. J. Gross and A. Neveu, Phys. Rev. D {\bf 10}, 3235 (1974).
\bibitem{L2}
Y. Nambu and G. Jona-Lasinio, Phys. Rev.  {\bf 124}, 246 (1961).
\bibitem{L3} 
G. 't Hooft, Nucl. Phys. B {\bf 72}, 461 (1974).
\bibitem{L4}
P. de Forcrand and U. Wenger, Proc. Sci. LAT2006, 152 (2006).
\bibitem{L5}
M. Wagner, Phys. Rev. D {\bf 76}. 076002 (2006).
\bibitem{L6}
T. Kojo, Y. Hidaka, L. McLerran, and R. D. Pisarski, Nucl. Phys. A {\bf 843}, 37 (2010).
\bibitem{L7}
T. Kojo, R. D. Pisarski, and A. M. Tsvelik, Phys. Rev. D {\bf 82}, 074015 (2010).
\bibitem{L8}
D. Nickel, Phys. Rev. D {\bf 80}, 074025 (2009).
\bibitem{L9}
M. Buballa and S. Carignano, Prog. Part. Nucl. Phys. {\bf 81}, 39 (2015).
\bibitem{L10}
J. Braun, F. Karbstein, S. Rechenberger, and D. Roscher, Phys. Rev. D {\bf 96}, 014032 (2016).
\bibitem{L11}
J. J. Lenz, L. Pannullo, M. Wagner, B. H. Wellegehausen, and A. Wipf, Phys. Rev. D {\bf 101}, 094512 (2020).
\bibitem{L12}
J. J. Lenz, L. Pannullo, M. Wagner, B. H. Wellegehausen, and A. Wipf, Phys. Rev. D {\bf 102}, 114501 (2020).
\bibitem{L13}
K. Horie and C. Nonaka, Proc. Sci. LAT2021, 150 (2021).
\bibitem{L14}
J. J. Lenz, M. Mandl, and A. Wipf, Phys. Rev. D {\bf 2022}, 034512 (2022).
\bibitem{L15}
O. Schnetz, M. Thies, and K. Urlichs, Ann. of Phys. {\bf 321}, 2604 (2006).
\bibitem{L16}
V. Sch\"on and M. Thies, {\it At the Frontier of Particle Physics: Handbook of QCD, Boris Ioffe Festschrift},
vol. 3, ed. M. Shifman (Singapore: World Scientific), ch. 33, p. 1945 (2001).
\bibitem{L17}
G. Basar, G. V. Dunne, and M. Thies, Phys. Rev. D {\bf 79}, 105012 (2009).
\bibitem{L18}
C. Boehmer, U. Fritsch, S. Kraus, and M. Thies, Phys. Rev. D {\bf 78}, 065043 (2008).
\bibitem{L19}
J. Braun, S. Finkbeiner, F. Karbstein, and D. Roscher, Phys. Rev. D {\bf 91}, 116006 (2015).
\bibitem{L20}
A. Koenigstein, L. Pannullo, S. Rechenberger, and M. Winstel, arXiv:2112.07024 [hep-ph].
\bibitem{L21}
Gordon Baym, {\it Lectures on Quantum Mechanics}, ch. 11,  Lecture Notes and Supplements in Physics, 
Benjamin/Cummings Publishing Company (1969).
\bibitem{L22}
A. Heinz, F. Giacosa, M. Wagner, and D. H. Rischke, Phys. Rev. D {\bf 93}, 014007 (2016).
\bibitem{L23}
M. Thies, Phys. Rev. D {\bf 101}, 014010 (2020).
\bibitem{L24}
M. Thies, Phys. Rev. D {\bf 101}, 074013 (2020).
\bibitem{L25}
C. Boehmer, M. Thies, and K. Urlichs, Phys. Rev. D {\bf 75}, 105017 (2007).
\bibitem{L26}
G. V. Dunne, J. Lopez-Sarrion, and K. Rao, Phys. Rev. D {\bf 66}, 025004 (2002).
\bibitem{L27}
I. Lindgren, J. Phys. B: Atom. Molec. Phys. {\bf 7}, 2441 (1974).
\bibitem{L28}
C. Boehmer, F. Karbstein, and M. Thies, Phys. Rev. D {\bf 77}, 125031 (2008).
\end{thebibliography}
\end{document}